\documentclass[useAMS,usenatbib,a4paper,referee]{mn2e}
\usepackage{graphicx}
\usepackage{amsmath}
\usepackage{txfonts}
\usepackage{mathrsfs}

\newcommand{\be}{\begin{equation}}
\newcommand{\ee}{\end{equation}}
\newcommand{\bea}{\begin{eqnarray}}
\newcommand{\eea}{\end{eqnarray}}

\title[Polarimetric Imaging of Sgr A*]{Polarimetric Imaging of Sgr A* in its Flaring State}
\author[Fulvio Melia, Maurizio Falanga, and Andrea Goldwurm]{Fulvio Melia,$^{1}$\thanks{Sir 
Thomas Lyle Fellow and Miegunyah Fellow. E-mail: melia@as.arizona.edu}
Maurizio Falanga,$^{2}$\thanks{E-mail: mfalanga@issibern.ch} and Andrea 
Goldwurm$^{3}$\thanks{E-mail: goldwurm@discovery.saclay.cea.fr}\\
$^{1}$Department of Physics, The Applied Math Program, and Department of Astronomy, 
The University of Arizona, AZ 85721, USA\\
$^{2}$International Space Science Institute (ISSI), Hallerstrasse 6, 3012 Bern, Switzerland\\
$^{3}$Service dAstrophysique (SAp), IRFU/DSM/CEA-Saclay, 91191 Gif-sur-Yvette Cedex, France;\\
Unit\'e mixte de recherche Astroparticule et Cosmologie, 10 rue Alice Domon et Leonie Duquet, 
75205 Paris, France}

\begin{document}

\date{}

\pagerange{\pageref{firstpage}--\pageref{lastpage}} \pubyear{2011}

\maketitle

\label{firstpage}

\begin{abstract}
The Galaxy's supermassive black hole, Sgr A*, produces an outburst of infrared radiation about once every
6 hours, sometimes accompanied by an even more energetic flurry of X-rays. It is rather clear now that
the NIR photons are produced by nonthermal synchrotron processes, but we still don't completely
understand where or why these flares originate, nor exactly how the X-rays are emitted. Circumstantial
evidence suggests that the power-law electrons radiating the infrared light may be partially cooled, allowing
for the possibility that their distribution should be more accurately described by a broken power law 
with a (``cooling break") transition frequency. In addition, the emission region (energized by an
as yet unidentified instability) appears to be rather compact, possibly restricted to the inner edge 
of the accretion disk. In that case, the X-ray outburst may itself be due to synchrotron processes 
by the most energetic particles in this population. In this paper, we examine several key features 
of this proposal, producing relativistically correct polarimetric images of Sgr A*'s NIR and X-ray 
flare emission, in order to determine (1) whether the measured NIR polarization fraction is 
consistent with this geometry, and (2) whether the predicted X-ray to NIR peak fluxes are 
confirmed by the currently available multi-wavelength observations. We also calculate the 
X-ray polarization fraction and position angle (relative to that of the NIR photons) in 
anticipation of such measurements in the coming years. We show that whereas the polarization 
fraction and position angle of the X-rays are similar to those of the NIR component for 
synchrotron-cooled emission, these quantities are measurably different when the X-rays 
emerge from a scattering medium. It is clear, therefore, that the development of X-ray 
polarimetry will represent a major new tool for studying the spacetime near supermassive 
black holes.

\end{abstract}


\section{Introduction}
Sagittarius A* (Sgr A*) is the radiative manisfestation of our galactic supermassive black hole. 
Discovered almost four decades ago (Balik \& Brown 1974), this intriguing radio source has been 
well studied at all wavelengths, generating interest among both theorists and observers (see 
Melia 2007 for a comprehensive review). A significant development with Sgr A* occurred in 2001, 
with the discovery of bright X-ray flares emanating from its vicinity (Baganoff et al. 2001). Lasting 
anywhere from 40 minutes to over 3 hours, these events are associated with enhanced X-ray emission, 
30--200 times greater than that of its quiescent state.

In the intervening years, many more X-ray flares have been observed by both {\it Chandra} and 
XMM-{\it Newton}, and analyzed extensively (see, e.g., Goldwurm et al. 2003, Porquet et al. 2003, B\'elanger
et al 2005, Porquet et al. 2008). Among the many intriguing features characterizing these events are temporal 
substructures as short as 200 s, indicating that the physical processes responsible for the outbursts 
occur within a mere 10--20 Schwarzschild radii of a $\sim 4\times 10^6\;M_\odot$ black hole (Sch\"odel et al. 
2002, Ghez et al. 2003).

Flaring events in Sgr A* have also been detected in the NIR (Genzel et al. 2003) and even at
mm and sub-mm wavelengths (see, e.g., Zhao et al. 2004, Trap et al. 2011). At least in the infrared, the flare spectra
appear to be power laws with variable linear polarization, suggesting optically thin synchrotron emission
as the likely mechanism producing the excess flux. But though the observational profile of these bursts 
continues to be refined, in tandem with significant theoretical discourse and modeling, we still don't 
completely understand where or why these flares originate, nor exactly how the X-rays are produced, 
or even if the variable mm and sub-mm emission is really indicative of the flares' spectral extension into 
the radio domain. One thing is certain, however---that no significant progress is likely to occur without 
multiwavelength observations and a broadband approach to the data analysis and modeling. For this 
reason, several groups have systematically organized large simultaneous observational campaigns, 
utilizing both ground-based and orbital platforms (see, e.g., B\'elanger et al. 2004, 2006, Yusef-Zadeh 
et al. 2006, Marrone et al. 2008, Yusef-Zadeh et al. 2009, Dodds-Eden et al. 2009, 
Kunneriath et al. 2010, Trap et al. 2011).    

The key results emerging from these (and other similar) studies are as follows (see Dodds-Eden et
al. 2009 for a more comprehensive compilation): (1) IR/NIR flares occur more frequently than X-ray
flares, typically $\sim 4$ times per day, compared with roughly 1 per day for the latter; (2) consequently,
every X-ray flare appears to be associated with an NIR flare, but not every NIR event is correlated
with an X-ray flare; additionally, (3) though X-ray and NIR flares occur simultaneously, with no significant delay,  
the former are typically shorter in duration;\footnote{Properties  (2) and (3) may be partially related by 
the fact that the flaring X-ray emission must reach a higher relative flux in order to be detectable above a steady 
background. Sgr A*'s quiescent X-ray flux appears to arise from thermal emission in the much larger accretion flow 
extending out to the capture, or Bondi-Hoyle, radius (Melia 1992, Ruffert \& Melia 1994, Yusef-Zadeh et al. 2000, 
Baganoff et al. 2003, Rockefeller et al. 2004). So the undiluted X-ray flare may not be as small compared
to its NIR counterpart as it appears above the background. But there are good reasons to believe
that the X-ray flare is indeed shorter than the NIR event, as discussed in Dodds-Eden et al. (2009).} 
(4) polarimetric measurements in 
the NIR show that the source is significantly polarized (by as much as $\sim 12\%-25\%$; see Eckart 
et al. 2006, Nishiyama et al. 2009; (5) the brightest flares have a constant spectral index $\alpha\sim 0.6$ 
between 3.8 and 1.6 $\mu$m, where the flux is given as $F_\nu\propto \nu^{-\alpha}$ (see, e.g., 
Hornstein et al. 2007). At lower intensities, there may also be a possible trend of spectral index 
with flux, though this is still in dispute.

Additional results emerge when we include observations at longer wavelengths, but our focus in
this {\it Letter} will be on the NIR and X-ray emission, so we will defer the more complete discussion
to a subsequent paper. As we shall see, three of the critical questions that must be resolved for
a complete understanding of the flare phenomenon in Sgr A* (at least in the NIR/X-ray)
are (1) how are the X-rays produced? (2) how do we understand Sgr A*'s polarized NIR 
flare emission?  and (3) can we predict the associated polarization fraction and position 
angle for the variable X-ray component, in anticipation of future polarimetric measurements 
at $\sim$ keV energies? Based on preliminary work (some of it by our group), we now 
suspect that the NIR/X-ray flares originate close to Sgr A*'s event horizon. As such, it will 
not be possible to carry out a meaningful simulation of the emission and polarization profiles 
without a general relativistic (GR) approach. Addressing these (and related) questions 
within the context of GR is therefore the principal goal of this {\it Letter}.

\section{Preliminary Considerations}
Sgr A*'s transient X-ray emission has variously been attributed to thermal processes
or invese Compton scattering---of either seed mm/sub-mm photons (by the relativistic
particles producing the IR/NIR synchrotron emission; see, e.g., Yusef-Zadeh et al. 
2006) or of the IR/NIR photons themselves as part of a synchrotron self-Compton (SSC) 
approach (see, e.g., Markoff et al. 2001, Liu \& Melia 2001, Eckart et al. 2004). It 
has also been suggested that both the NIR and the X-ray components may be due to the 
same synchrotron process (see, e.g., Yuan et al. 2004).

But the analysis of several bright flares detected over the past few years has all 
but ruled out these earlier proposals (see Dodds-Eden et al. 2009, Trap et al. 2011). 
Single component synchrotron models (with a particle distribution $dN(\gamma)
\propto {\gamma}^{-p}\,d\gamma$) are problematic because the high-energy
electrons required to generate X-ray synchrotron emission have very short
cooling times (much shorter than a typical X-ray flare duration). Thus, even a 
continuous injection to replenish the energetic population cannot sustain the 
same power-law index $p$ at both low and high energies.

If instead the X-rays are submm photons inverse-Compton scattered
by the electrons producing the NIR emission, the largest permissible size
of the quiescent radio-producing region is about $0.27\,R_S$ (Dodds-Eden
et al. 2009, Trap et al. 2011), far smaller than the measured FWHM size ($\approx
3.7\,R_S$; Doeleman et al. 2008) of Sgr A* at 1.3 mm. This uncomfortably tight
restriction (see also Liu \& Melia 2001) is compounded by several other difficulties, 
but even the size issue on its own already rules out the ``external" Compton 
scattering scenario.

Attempts at fitting an SSC spectrum to the combined NIR/X-ray data are equally
problematic because such a model requires low electron energies (with $\gamma
\sim$ 10--15) and unrealistically strong magnetic fields ($B>1,000$ G) and very 
large particle densities (much larger than those inferred for the quiescent emission).

We are thus left with the following rather tightly constrained indicators. During a typical
flare, there is little if any detectable emission at $11.88\;\mu$m, implying that
the flare emission spectrum (characterized by the power density $\nu F_\nu$)
must rise from the MIR towards the NIR. This is consistent with the spectral
index $\alpha\sim 0.6$ described above. It is clear, therefore, that the electron
population producing the $L^\prime$-band flare has a different distribution of
energies than that associated with the submm bump (see, e.g., Melia 2007),
so a NIR flare cannot simply be a small change in the overall properties of
the steady radio-submm emitting region.

However, it is unrealistic to expect a power-law particle distribution such as
this to maintain the same power-law index $p$ at all energies. Synchrotron energy 
losses, not to mention the escape time from the acceleration region, both depend on 
the particle energy (see, e.g., Liu et al. 2006). 
It is well known (see, e.g., Pacholczyk 1970) that while
energetic electrons are injected continuously into the system by the acceleration
process, the emitted steady-state photon spectrum has an index $\alpha=(3-p)/2$
(with $p$ the particle index) up to a ``cooling break" frequency $\nu_b$, 
steepening to an index $\alpha=(2-p)/2$ (corresponding to a particle index 
$p+1$) above it. The cooling break corresponds to the electron energy (or
Lorentz factor $\gamma_b$) at which the escape time $\tau_{esc}$ from 
the system is equal to the (synchrotron) cooling time $\tau_{cool}$.

The cooling timescale is given as (Pacholczyk 1970)
\begin{equation}
\tau_{cool}\approx 8\left({B\over 30\;{\rm G}}\right)^{-3/2}\left({\nu
\over 10^{14}\;{\rm Hz}}\right)^{-1/2}\quad{\rm min}\;,
\end{equation}
but the escape timescale depends on several factors, including $B$ and
the ambient particle density $n_e$. Nonetheless, detailed calculations of 
the acceleration process (see Nayakshin \& Melia 1998,  Liu et al. 2004, 2006) 
indicate that for a broad range of conditions, $\tau_{esc}$ corresponds 
roughly to 3 times the light transit time in the system. (For an analogous,
though much smaller system than Sgr A*, see also Misra \& Melia 1993.)
Thus, if the emission region is near the marginally stable orbit (see below), 
we would expect
\begin{equation}
\tau_{esc}\sim 3{6GM\over c^3}\approx 6\;{\rm min}\;,
\end{equation}
for a black hole mass $4\times 10^6\;M_\odot$.
In the overall spectrum, the cooling break is therefore expected to occur at a frequency
\begin{equation}
\nu_b\approx \left(2\times 10^{14}\;{\rm Hz}\right)\left({B\over 30\;
{\rm G}}\right)^{-3}\;,
\end{equation}
i.e., somewhere above the NIR component and well below the X-ray, as long
as the magnetic field intensity is of order 30 G, essentially the value required
to produce Sgr A*'s quiscent emission.

For the simulations reported in this {\it Letter}, we will therefore assume that
the relativistic particle distribution producing the NIR/X-ray flare has the 
following form:
\begin{equation}
dN(\gamma)=N_0{
\begin{cases}
\text{$\gamma^{-p}\,d\gamma$} & \text{$\gamma < \gamma_b$} \\
\text{$\gamma^{-(p+1)}\,d\gamma$} & \text{$\gamma > \gamma_b$}
\end{cases}}\;,
\end{equation}
where the choice of $p$, $\gamma_b$ (or, equivalently, $B$ in Equation~3), and the 
normalization constant $N_0$, are all based on earlier fits to the data (see, e.g.,
Dodds-Eden et al. 2009, Trap et al. 2011). 

An additional constraint on the flaring region is provided by
the temporal substructure seen in typical bursts, particularly in the NIR. For
example, one sees $L^\prime$-band flux variations up to $\sim 30\%$ of the 
peak value on a timescale of only $\sim 20$ minutes. Taken at face value,
these fluctuations could be telling us that most of the burst activity is
occurring near the marginally stable orbit surrounding a $\sim 4\times 10^6\;M_\odot$
black hole (see, e.g., Melia et al. 2001a). However, light-travel arguments
constrain the size of the emitting region even further when one uses the
shortest time-scale variations seen, e.g., in the 4 April, 2007 burst reported
by Dodds-Eden et al. (2009). There, in the $L^\prime$ lightcurve, very rapid
changes in flux (by factors of 120--170 $\%$, with a statistical significance
$>3\sigma$) were observed within a timescale $\Delta t<47$ seconds. 

Thus, a large fraction of the emitting plasma must be confined to a region 
$<c\Delta t$ which, for a $\sim 4\times 10^6\;M_\odot$ black hole, corresponds 
to about $1.2\,R_S$, where $R_S\equiv 2GM/c^2$ is the Schwarzschild 
radius. We may be witnessing the emergence of a very confined region, such 
as a magnetic flux tube breaking out above the disk, or perhaps a ring-like 
``hot" region near the marginally stable orbit. The latter scenario, in particular,
would be expected on the basis of an instability in the disk, such as the
Rossby-wave instability studied in detail by Tagger \& Melia (2006) and
Falanga et al. (2007).

Our goal in this {\it Letter} is to ascertain whether these observed
NIR/X-ray flare characteristics can be explained in terms of the geometry 
and the particle distributions described above, taking into account all of
the essential general relativistic effects, such as light bending and area 
amplification. Unfortunately, what is not known precisely is the fraction of 
transient X-ray emission hidden below Sgr A*'s quiescent high-energy luminosity 
(which presumably originates from a different location, as discussed earlier). 
As such, our calculation of the X-ray to NIR intensity ratio 
necessarily provides only an upper limit to the observed value, but because we
here work principally with a single particle population (Equation~4), even
this upper limit is very probative. (As our understanding of the conditions in
the ambient medium surrounding Sgr A* improves, the uncertainty in this
fraction will not doubt diminish; see, e.g., Crocker et al. 2010.)

This is because in addition to the total flux densities themselves, we
will also carefully calculate the polarization fractions expected in
this scenario for both the NIR and X-ray components. The discovery
of significant linear polarization in the NIR (see, e.g., Eckart et al.
2006) was an important milestone in the study of these flares
because it unambiguously confirmed the signature of nonthermal
synchrotron radiation. But we still don't know if this measured
{\it degree} of polarization is consistent with the geometry and 
particle distribution suggested by the other flare characteristics.
Our simulations here should help us resolve this question.

And an equally important question concerns the polarized fraction
of X-rays. Of course, we do not yet have data relevant to this issue. 
However, it is not unreasonable to expect the X-ray polarization 
fraction to differ from that of the NIR (since the polarized emissivities 
are energy dependent; see below), and to differ considerably between 
two possible origins, (i) synchrotron radiation due to a cooled power-law 
particle distribution, or (ii) synchrotron-self-Comptonization. Thus, that
portion of our simulation involving X-rays is motivated not only by our 
desire to provide a firm prediction of the X-ray polarization fraction
for comparison with future observations, but also to demonstrate
how the degree of X-ray polarization relative to that in the NIR
differs depending on the mechanism producing the X-rays. In
so doing, we hope to establish the viability of an important 
diagnostic for examining whether or not the X-rays are indeed 
produced by a cooled power-law population of electrons, and 
whether they originate from a small region near the marginally 
stable orbit.
  
\section{Polarimetric Imaging of the NIR/X-ray Emitting Region}
Our group has been developing methods to produce synthetic images of Sgr A*
using general-relativistic ray-tracing codes since the 1990's (Hollywood \& Melia
1997, Falcke et al. 2000, Bromley et al. 2001, Falanga et al. 2007). Other
investigators have by now successfully produced their own algorithms for carrying 
out similar ray-traced simulations, producing a rich ensemble of complementary 
results (see, e.g., Broderick \& Loeb 2005, Zamaninasab et al. 2011).
In a broader context, the transfer of polarized radiation through black-hole
spacetimes has also been explored by Dovciak et al. (2004), Noble et al.
(2007), Dolence et al. (2009), Dexter et al. (2009), Davis et al. (2009),
Schnittman \& Krolik (2010), and Shchenbakov \& Huang (2011).
Spurred by the evident need to carry out a comprehensive broadband
investigation of Sgr A*'s flaring state, in which the radiation we observe
at Earth may be polarized, but which may also include a Comptonized 
component  (possibly the X-rays) originating from locations other than 
those where the primary (or seed) photons are emitted, we have recently 
developed a more comprehensive code called POLLUX (POLarized
LUX). This algorithm (more fully described in Falanga et al. 2011) 
incorporates several indispensible physical attributes associated with the
emission and propagation of a multi-component spectrum through the
Schwarzschild spacetime surrounding a non-spinning black hole. Often,
the presence of a relativistic jet is taken to be the signature of a high
spin rate. The fact that no such structure has ever been detected
in Sgr A* suggests that this object may be spinning  slowly, if at all
(but see also Melia et al. 2002, Liu \& Melia 2002). Thus, in order to 
keep the simulations as straightforward as possible, we have chosen 
to work with the Schwarzschild metric in this instance. 

At the inner edge of the disk, where the NIR and X-ray photons are
presumably being emitted, the gas is circularized and settled into 
Keplerian motion, where a magnetohydrodynamic dynamo produces
a predominantly azimuthal magnetic field ${\bf B}$$\approx$$ B\hat\phi$ 
(Hawley, Gammie \& Balbus 1996). Overcoming the rate of field 
destruction in the differentially rotating portion of the inflow, the 
field reaches a saturated intensity since the dynamo timescale is 
shorter than the dissipation timescale in this region (Melia et al. 2001b).
However, it is well-known by now that the disk surrounding Sgr A*
cannot be a large standard disk, which would produce a detectable 
quiescent-state IR emission (see, e.g., Falcke \& Melia 1997). Thus,
all of these processes, even in the quiescent state, must be restricted
to a compact region no bigger than 10--20 Schwarzschild radii from
the black hole.

For the incipient emission at NIR wavelengths, we track both the extraordinary
(perpendicular to {\bf B}) and ordinary (parallel to {\bf B}) waves, whose 
emissivities may be written (see, e.g., Rybicki \& Lightman 1985):
\begin{equation}
\epsilon^e(\omega)=K\left[{1\over 1+p}\Gamma\left({p\over 4}+{17\over 12}\right)
+{1\over 4}\Gamma\left({p\over 4}+{7\over 12}\right)\right]\omega^{(1-p)/2}\;,
\end{equation}
and
\begin{equation}
\epsilon^o(\omega)=K\left[{1\over 1+p}\Gamma\left({p\over 4}+{17\over 12}\right)
-{1\over 4}\Gamma\left({p\over 4}+{7\over 12}\right)\right]\omega^{(1-p)/2}\;,
\end{equation}
where
\begin{equation}
K\equiv {3^{p/2}e^{(5+p)/2}N_0B^{(1+p)/2}\over 16\pi^2 {m_e}^{(1+p)/2}c^{(3+p)/2}}
\Gamma\left({p\over 4}-{1\over 12}\right)\left(\sin\theta_e\right)^{(3-p)/2}\;.
\end{equation}
In these expressions, $\omega$ is the angular frequency in the comoving frame,
$\Gamma$ is the Gamma function, and $\theta_e$ is the angle between the local 
magnetic field vector and the outwardly pointing ray (which is also very nearly the
pitch angle since for relativistic particles most of the emission is projected into a very tiny
cone about the particle's velocity). When the X-rays are produced predominantly by the
cooled electrons in the broken power law (Equation~4), as opposed to inverse Compton
scattering, which we also consider in this {\it Letter}, their extraordinary and ordinary
emissivities are given by expressions analogous to these, except with $p\rightarrow
p+1$.

The overall specific intensity ${I_\nu}^{e,o}$ observed at infinity is an integration
of the emissivity $\epsilon^{e,o}$ over the path length along geodesics, though most
of the contribution arises in the geometrically thin disk, close to the marginally stable
orbit. However, a novel feature of our code is that we also allow the incipient radiation
to scatter with the ambient plasma, either thermal or nonthermal, as the case may be, 
as it propagates along the geodesics towards the observer. Throughout this process,
we preserve (or update as needed) the photon energy, the overall intensity, and the 
polarization fraction and position angle. As we shall see, this allows us to form polarimetric 
images at infinity of both the {\it direct} emission from the transiently energized inner region 
of the disk and the light {\it scattered} above or beyond this location.

It is straightforward to relate the specific intensity of each polarization separately in the
emitter and detector frames from the relativistic invariant $I_\nu/\nu^3$, and the degree
of polarization is itself an invariant, though the position vector may rotate along
geodesics. Likewise, the frequency $\nu(r)$ at radius $r$ from the central object, 
is known precisely in terms of the frequency $\nu$ in the frame comoving with the 
emitter, and the lapse function, $\Lambda$, which includes both the effects of 
gravitational time dilation (between the emitter's radius $r_e$ and the pertinent
radius $r$) and the Doppler shift associated with motion of the emitting particles
at $r_e$:
\begin{equation}
\Lambda=(1-R_s/r_e)^{1/2}(1-R_s/r)^{-1/2}(1-\beta^2)^{1/2}(1-\beta\cos\theta_e)^{-1/2}\;,
\end{equation}
where $\beta=v/c$ in terms of the emitting particle's velocity $v$, and $\theta_e$ is the 
aforementioned angle between the local magnetic field vector and the outwardly pointing
ray, since {\bf B} and {\bf v} are parallel in the assumed Keplerian geometry. At radius
$r$, we have $\nu(r)=\Lambda \nu$, and the intensity then follows from the relativistic
invariant. For the image of the direct emission viewed at infinity, we simply set
$r\rightarrow\infty$, while for the scattered light, $r$ becomes the radius $r_{sc}$
at which the scattering has occurred, and an additional ``scattered" lapse function
$\Lambda_{sc}=(1-R_s/r_{sc})^{1/2}$  is introduced to incorporate the effects of
gravitational time dilation between $r_{sc}$ and infinity. (Note that we are not including
an additional Doppler shift factor into the expression for $\Lambda_{sc}$ because the
frequency shift associated with the scattering itself is already included in the evaluation
of the Comptonized photon's energy.)

The position angle of polarized light may be calculated from a
relativistic invariant related to the parallel transport of a polarization vector along the
null rays (see, e.g., Connors \& Stark 1977). We perform the parallel transport operation
quite directly by defining a reference vector at the detector and numerically propagating
it along with the null ray itself. This allows us to consistently map a position angle from
one frame to the next, an essential feature for calculating radiative transfer for the
ordinary and extraordinary waves as they propagate around the black hole.

Scattering introduces a significant complication to this process because it not only
redirects the photon's trajectory, but also creates its own polarization. We follow
the method described in Connors et al. (1980), based on the use of normalized
Stokes parameters
\begin{equation}
X_s={Q\over I},\qquad Y_s={U\over I}\;,
\end{equation}
where $I$ is the intensity and $Q$, $U$ are the Stokes parameters determining
the linear polarization. The Stokes parameters $X_s^\prime$ and $Y_s^\prime$
of the scattered beam can be determined from $X_s$ and $Y_s$, once the
scattered photon's new direction is known and a new set of reference axes
have been defined from the wavevectors before and after scattering. We
refer the reader to this well-written paper for further details.

\begin{figure}
\center{\includegraphics[scale=0.5,angle=0]{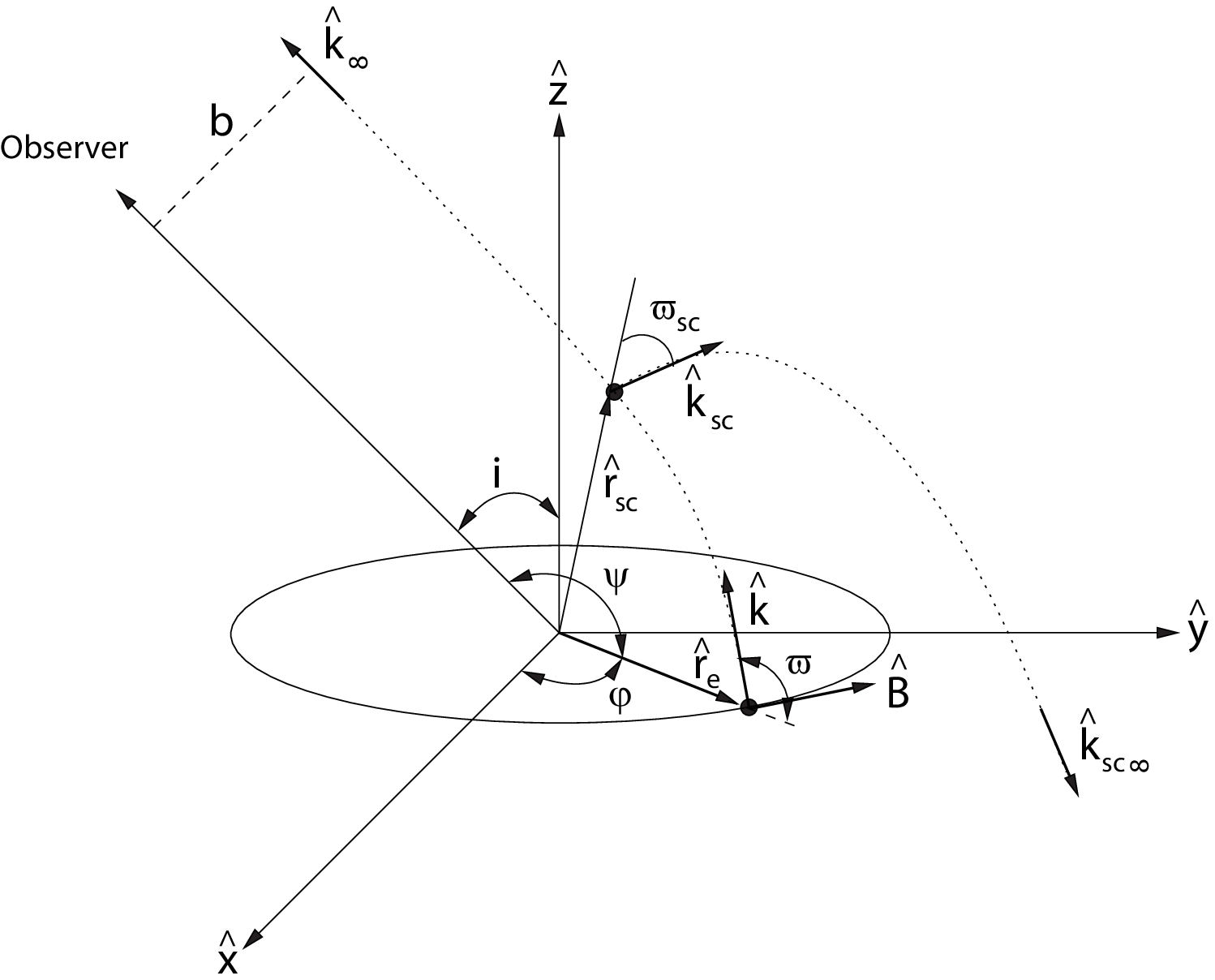}
\caption{Schematic diagram showing the relevant angles and unit vectors
defining our geometry. The unit vector $\hat{\rm k}$ denotes the photon's direction,
$\hat{\rm B}$ points in the direction of the magnetic field, and quantities
with subscript ``sc" correspond to the scattered photon. Other symbols 
are defined in the text. For illustrative purposes, two sample photon trajectories
are shown: first, the unperturbed geodesic emanating from the disk; 
second, the modified trajectory resulting from a scattering event at 
${\bf r}_{sc}$.}}
\end{figure}

The observer is located at infinity with viewing angle $i$ relative to the $z$-axis
in the non-rotating frame (see figure~1). The deflection angle of a photon emitted by plasma
in the transiently energized inner region of the disk is $\psi$, where $\cos\psi=\cos i\cos\phi$, 
$\phi$ being the azimuthal angle of the emission point relative to the reference 
$x$-axis. These emitted photons are deflected by the black hole and intersect the 
observer's detector plane at infinity. (Note that this same geometry is valid for
the deflection of scattered photons as well.) The distance
between the line-of-sight and the point at which the photon reaches the detector
is defined as the impact parameter $b$. Using this geometry, the deflection
angle of the photon's trajectory may be obtained with the light-bending
relation between $\varpi$ (the angle between the emission direction
of the photon and the direction from the center of the black hole to the
location of the emitter) and $\psi$, from the geodesic equation (see
Beloborodov 2002)
\begin{equation}
\cos\psi=1-{1-\cos\varpi\over 1-R_s/r}\;.
\end{equation}
This procedure yields the impact parameter $b=r_e(1-R_s/r_e)^{-1}\sin\varpi$ of the
photons in terms of the emitting radius $r_e$. A detailed description of this geometry
is provided in, e.g., Luminet (1979), and a more complete accounting of POLLUX,
together with several other sample applications, will appear in Falanga et al. (2011).

%\begin{figure}
%\center{\includegraphics[scale=0.55,angle=0]{fig1.eps}
%\caption{\footnotesize }
%\end{figure}

\section{Simulations}
Given the above considerations---some from prior fitting to the data (e.g.,
Dodds-Eden et al. 2009, Trap et al. 2011), others from earlier theoretical
work (e.g., Melia 2007)---we have selected the following physical environment
to model with POLLUX, the results of which are summarized in the next two 
subsections. We assume a radiating ring (energized by an as yet unidentified
instability), $1\;R_S$ wide and inner radius at $3\;R_S$. The non-thermal particle 
distribution is described by the synchrotron-cooled power law in Equation~(4), with 
an ambient azimuthal magnetic field ${\bf B}=(10\;{\rm G})\,\hat{\phi}$. The break 
frequency (Equation~3) is then $5.4\times 10^{15}$ Hz. Aside from the clear observational 
motivation for this geometry, there are also compelling theoretical reasons for expecting
this type of disk instability, as discussed more extensively 
in Tagger \& Melia (2006) and Falanga et al. (2007). Further, to examine how
the polarimetric image of this instability compares with one in which scattering
is important, we also assume (and report in subsection 4.2 below the impact of) 
a straightforward scattering halo, with uniform electron density $n_{sc}=10^{11}$ 
cm$^{-3}$ and radius $15\;R_S$, centered on the black hole. Again, for specificity, 
we take these particles to be thermal, with temperature $T_{sc}=10^9$ K. 

\begin{figure}
\center{\includegraphics[scale=0.84,angle=0]{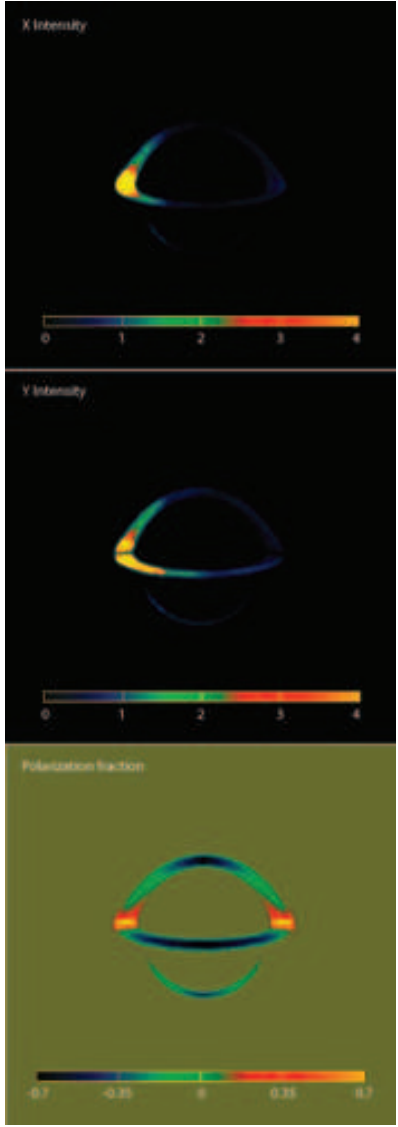}
\caption{Polarimetric images of a radiating ring, with width $1\;R_S$ and inner
radius $3\;R_S$, producing a cooled synchrotron spectrum with a break frequency
at $5.4\times 10^{15}$ Hz. The images shown here correspond to the frequency
range $1.6\times 10^{14}$ Hz to $2.8\times 10^{14}$ Hz, as viewed by an observer
at an inclination angle of $75-80$ degrees relative to the disk's symmetry axis.
{\it Top panel}: Light polarized in the horizontal (or $x$) direction;
{\it Middle panel}: Light polarized in the vertical (or $y$) direction;
{\it Lower panel}: Polarization fraction defined in Equation~(11).}}
\end{figure}

\subsection{Single Synchrotron-cooled Power Law}
Figure~2 gives a sample of the results from this simulation. In all of these images,
the vertical (or $y$) coordinate is parallel to the disk's symmetry axis. The top panel 
shows the intensity $I_x$ of light polarized in the horizontal (or $x$) direction on a linear color
scale.\footnote{Note that modeling the particle acceleration itself is beyond the
scope of this paper, so we are not attempting here to reproduce the measured intensity,
though the relative scaling---and the {\it spectrum}---are physically correct.} The
corresponding intensity $I_y$ of light polarized in the $y$ direction appears in the middle panel.
The bottom panel is an image (using the same spatial scale as the first two) of the
polarization fraction, defined as
\begin{equation}
\Pi\equiv {F_x-F_y\over F_x+F_y}\;,
\end{equation}
where $F_x$ and $F_y$ are the fluxes of polarized light calculated in the $x$ 
and $y$ directions, respectively, by integrating $I_x$ and $I_y$ over
their respective solid angles, with all the usual gravitational effects---e.g.,
area amplification---taken into account.

The interpretation of these results is rather straightforward (see also
Bromley et al. 2001). As we discuss in \S~3 above, the dominant magnetic field
component in this region is expected to be azimuthal. The extraordinary emissivity
dominates over the ordinary, so most of the $y$-polarized light comes from the front
and rear of the disk, whereas most of the $x$-polarized light comes from the sides.
In addition, the disk is assumed to rotate counter-clockwise in this geometry, so the
left is blue-shifted, whereas the right is red-shifted. All the general relativistic
and emissivity effects together conspire to produce a shift to the left of
$I_x$ relative to $I_y$. However, the polarized fraction is not affected by 
redshift, so the bottom image is symmetric about the vertical axis, and more 
clearly demonstrates the spatial dependence of the extraordinary versus ordinary 
emissivities. In principle, the polarization fraction varies over a broad range in 
values, from $-0.7$ (the negative sign meaning that $I_y$ is dominant over $I_x$)
to $0.7$ (when $I_x$ dominates over $I_y$). What we see, however, is an integration
of the intensity over the whole image, yielding a single polarized flux, so the
net polarization fraction is never quite this high (see figure~4 below).

The polarimetric images at other energies, say $2-10$ keV, are similar to these,
and will be reported as part of a more complete survey of results in Falanga et
al. (2011). But we will continue our discussion of these characteristics in
conjunction with figure~4 below.

\begin{figure}
\center{\includegraphics[scale=0.88,angle=0]{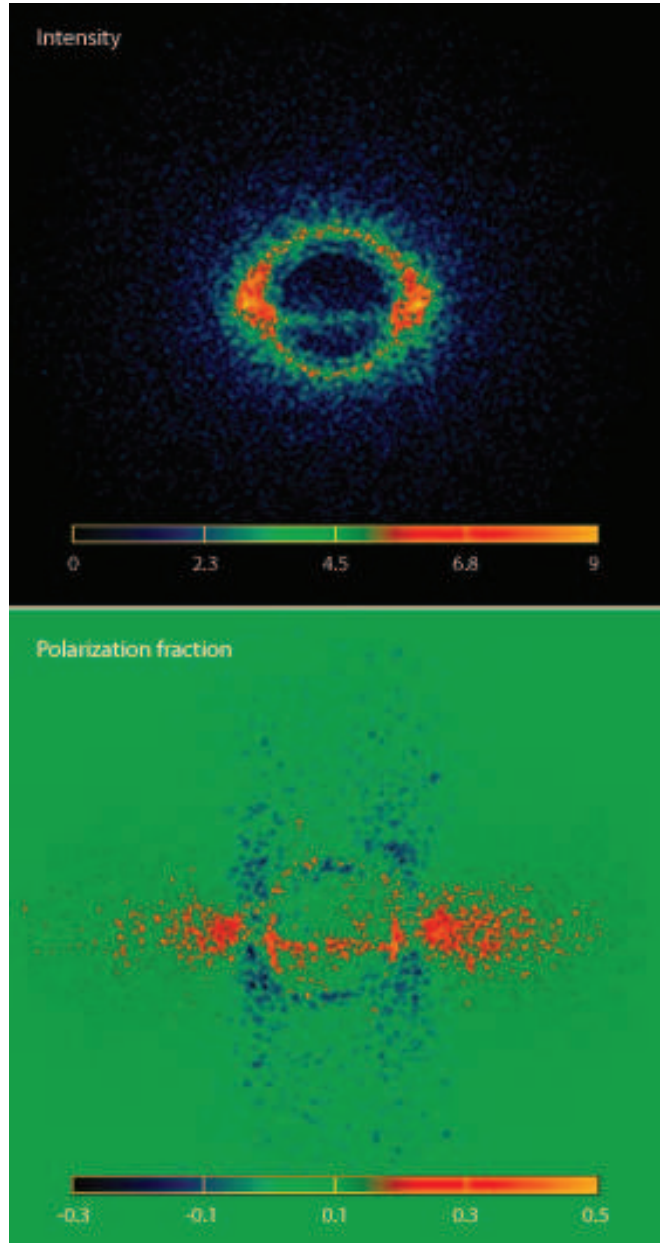}
\caption{Same as figure~2, except here for the total intensity $I=I_x+I_y$,
imaged in the frequency range $5\times 10^{17}$ Hz to $1.1\times 10^{18}$ Hz,
passing through a uniform scattering halo with radius $15\;R_S$ centered on
the black hole. {\it Upper panel}: The logarithm of total intensity;
{\it Lower panel}: Polarization fraction.}}
\end{figure}

\subsection{Synchrotron NIR Plus Comptonized X-rays}
Before doing so, however, let us examine the impact on our results of a 
scattering halo. Let us assume now that, in addition to the synchrotron
emitting ring described above, we also have a scattering halo surrounding the
black hole and inner disk region. We are not necessarily espousing the view
that this is a viable model for the NIR/X-ray flare characteristics observed
in recent years. If Comptonization is important in producing the X-rays,
the incipient emission need not be that due to a synchrotron-cooled particle
distribution. Our main goal here is to demonstrate how the images and
polarization fraction (and possibly the position angle) are altered by
scattering from those corresponding to the {\it direct} images.

Scattering clearly broadens the image, by a degree dependent on the density
$n_{sc}$ of scattering particles. In this simulation, we have chosen a value
$n_{sc}=10^{11}$ cm$^{-3}$, corresponding to a photon mean free path roughly
equal to the size of the halo. Thus, virtually all of the photons emitted
by the inner ring scatter at least once on their way out.

\begin{figure}
\center{\includegraphics[scale=0.78,angle=0]{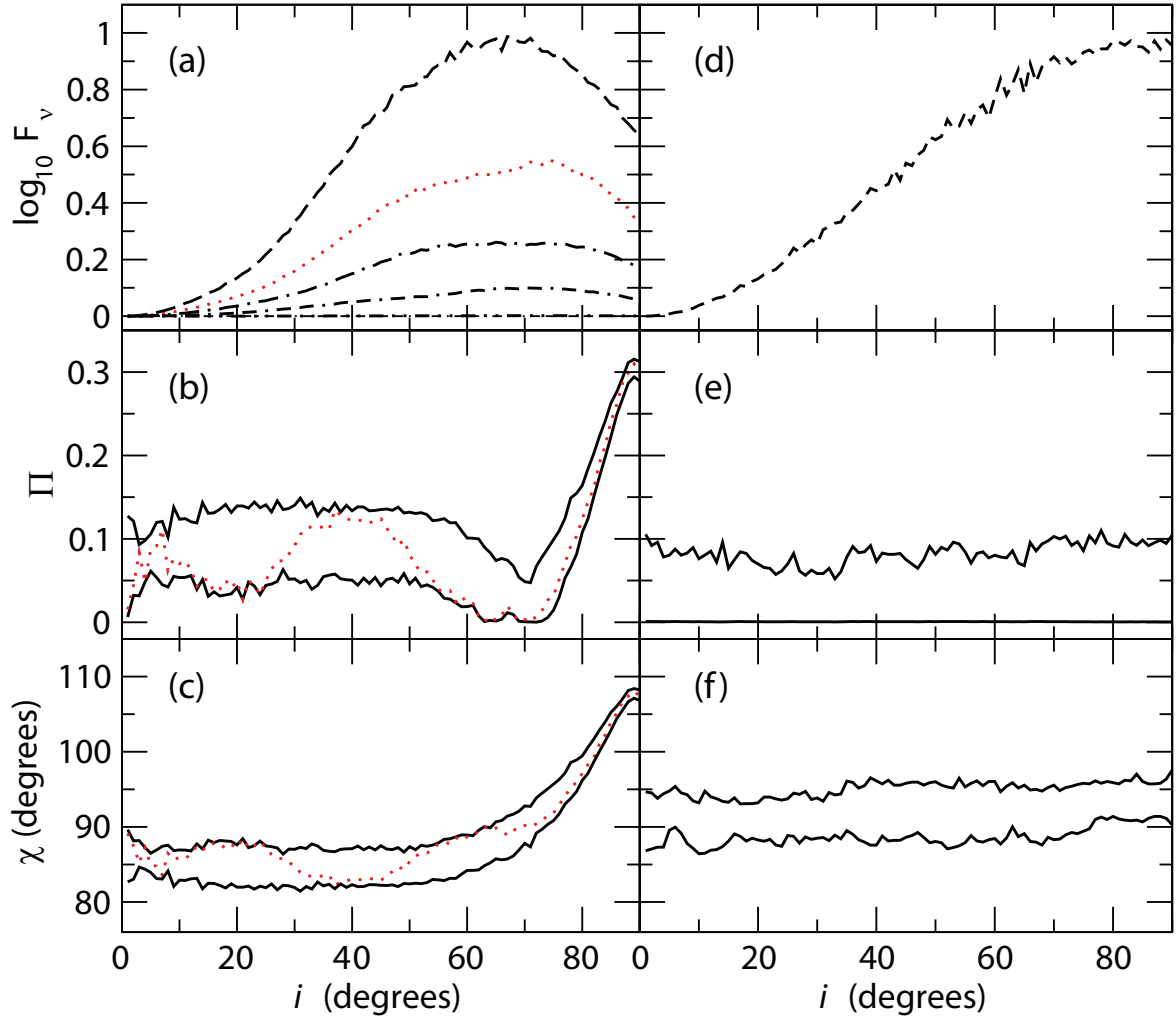}
\caption{{\it Panel (a)}: total flux density as a function of inclination
angle, for various frequency ranges. From top to bottom, the 5 curves
correspond to $(5\times 10^{13},9\times 10^{13})$ Hz, 
$(1.6\times 10^{14},2.9\times 10^{14})$ Hz,
$(5\times 10^{14},9\times 10^{14})$ Hz,
$(1.6\times 10^{15},2.9\times 10^{15})$ Hz, and $(5\times 10^{17},9\times 10^{17})$ Hz.
{\it Panel (b)}: Solid curves showing the maximum and minimum polarizations for any
of the frequencies, $(2.9\times 10^{13}, 2.9\times 10^{19})$ Hz, included in our 
simulation. By way of example, the dotted curve in this panel corresponds to
the polarization fraction as a function of inclination angle for the frequency
range $(1.6\times 10^{14}, 2.9\times 10^{14})$ Hz. The rise in $\Pi$ as
$i\rightarrow 40^\circ$ is due to the dominance of the Doppler-shifted emission
from the sides of the disk, whereas the decline in $\Pi$ as $i$ increases further
is due to the increasing influence of the emission from the front and rear of the
disk. The position angle $\chi$ changes in concernt to reflect these varying
contributions. {\it Panel (c)}: The range of
position angle $\chi$ (defined in Equation~12) as a function of inclination angle,
for the same frequencies as in the previous two panels.
{\it Panel (d)}: total flux density of light scattered into the frequency range
$(1\times 10^{14},2.5\times 10^{14})$ Hz. {\it Panel (e)}: Same as Panel (b),
except now for the scattered light (see figure~3). Here, the lower limit on $\Pi$
is zero. {\it Panel (f)}: Same as Panel (c), except for the scattered light.}}
\end{figure}

But scattering does more than this. It is well known that radiation passing
through a scattering medium acquires a linear polarization fraction dependent 
on the inclination angle relative to the surface of the scattering
medium (see, e.g., Conors et al. 1980). Thus, light originally polarized
at the point of emission, may become more or less polarized, depending
on its trajectory through to the observer. We see the tell-tale signature
of this effect in the lower panel of figure~3. Notice, e.g., that portions of
the image polarized in the vertical direction (in blue) now extend much farther 
above and below the disk. And we shall see in figure~4 below that an important
consequence of scattering is a general dilution of otherwise more strongly polarized
X-ray emission.

\section{Key Results and Conclusions}
Some of the most important results of our simulations are summarized in
figure~4. With these, we can now begin to answer some of the questions
posed in the introduction. First of all, a broken power-law distribution of
particles emitting synchrotron radiation from an energized ring at the marginally
stable orbit produces a polarization fraction at infinity of up to $\sim 15\%$
(even more for inclination angles $>70^\circ$) when all the relevant general
relativistic effects are taken into account. Polarimetric measurements in the NIR 
reveal polarization fractions as high as $12\%-25\%$. The geometry we have 
adopted here can therefore account for these observations without the need 
to induce linear polarization using more complicated procedures.

Detailed fits to the observations led to the idea that the X-rays may be due
primarily to the same population of electrons producing the NIR emission, though
synchrotron-cooled at higher energies. The results of our simulations show that
this remains a viable model even when all of the relativistic effects are taken
into account. Null geodesics (and parallel transport) are not energy dependent, so
the polarization properties of the X-ray component echo those of the NIR. Therefore,
a prediction of this model is that both the polarization fraction and position
angle $\chi$ of the X-rays should be similar to those of the infrared. Panel (c) in
figure~4 shows that $\chi$, defined by the expression
\begin{equation}
\chi\equiv \cos^{-1}\left({F_x-F_y\over F_x+F_y}\right)\;,
\end{equation}
lies somewhere between $85^\circ$ and $90^\circ$ for most inclination
angles, except above $\sim 70^\circ$ where it can reach $\sim 110^\circ$ or
more. These values of position angle are consistent with the
fact that $F_y$ dominates over $F_x$, as one would expect when the extraordinary
component is much greater than the ordinary. The increase in $\chi$ with $i$ is due 
to the increasing influence of redshift seen, e.g., in figure~2, which moves the
core of the observed emission towards the left in these images. 

Notice, however, that a different picture emerges for the X-ray characteristics 
when scattering is important. Panel (e) in figure~4 shows that the polarization fraction
is smaller (typically smaller than $10\%$) when X-rays emerge from scattering. For
many inclination angles, $\Pi$ is actually closer to $0\%$. In addition, the position
angle differs from that in Panel (c), by anywhere from $\sim 5^\circ$ to as much as
$20^\circ$, depending on the inclination angle.

Of course, the capability to measure $\Pi$ and $\chi$ for X-rays may be many 
years away, but that technology is being developed. For example, several X-ray 
polarimetry missions have already been studied and proposed to open up this very 
promising field of research over the next few years. The most advanced project 
today---the Gravity and Extreme Magnetism (GEM) mission (Swank et al. 2010), is
planned for launch in 2014, with the primary goal of studying the polarimetric properties
of the soft X-ray emission from bright black holes and neutron stars. Unfortunately,
the GEM instruments will not be sensitive enough to study a relatively weak source,
such as Sgr A*. Indeed, for fluxes less than 1 milliCrab (the intensity reached by
Sgr A* at the peak of the brightest flares), GEM would need a net observing time 
of more than 100 ks to detect a polarization fraction of 10\% or lower. But Sgr A*'s 
flares never last longer than 3 hr, so it would not be feasible to obtain 
such a deep exposure. Moreover GEM, and even other future polarimetry projects,
such as POLARIX (Costa et al. 2010), could not provide angular resolutions better 
than $\sim 1^\prime$, making it very difficult to identify the origin of a polarized 
signal from a crowded region, such as we have at the galactic center. Nonetheless,
this field is only in its infancy. More promising missions, with better sensitivity and
spatial resolution, will no doubt follow in the years to come. And our results 
demonstrate the power of future simultaneous polarimetric observations in both 
the NIR and X-ray portions of the spectrum.  

\section*{Acknowledgments}
This research was supported by NASA grant NNX08AX34G at the University of Arizona.
Partial support was also provided by ONR grant N00014-09-C-0032. In addition, FM 
is grateful to Amherst College for its support through a John Woodruff Simpson 
Lectureship. We also acknowledge the International Space Science Institute (ISSI)
in Bern, where a large portion of this work was carried out.

\end{document}